\documentclass{kapproc}
\usepackage{psfig}
\begin{document}
\articletitle[]{Dispersion in modeled abundances}

\author{Juan Jos\'{e} Baz\'{a}n, Mercedes Moll\'{a}} 

\affil{Dpto. de F\'{\i}sica Te\'{o}rica.Universidad Aut\'{o}noma
de Madrid, 28049 Cantoblanco, Madrid.}

\author{Miguel  Cervi\~{n}o}
\affil{IAA/CSIC, Apartado 3004, 18080, Granada.}

\section{Introduction}

Chemical abundance data for the Galaxy shows a wide dispersion.  We
try to check if this dispersion may also be found with chemical
evolution models as an effect of variations of an Initial Mass
Function (IMF) which follows a Poisson's distribution.

We will use two different methods: a) an analysis of the model
internal error and b) a statistic study on the effects of the IMF
discretization, based on montecarlo models. Both methods have been used
in a chemical evolution model applied to the Solar Neighborhood. 
Preliminary results relative to O and N are shown.

\section{The Initial Mass Function}

We will try to justify the data dispersion by the possible variations
in the model results which could be produced by the behavior of the
IMF, whose properties are described by a probability distribution
function:

If we assume that N$_{tot}$ stars are observed in a mass range
[m$_{0}$, m$_{max}$], the mass m$_{i}$ of the i-th star is a random
variable whose probability distribution function is given by the
stellar initial mass function: $\Phi (m_i )$, whose integration along
the whole mass range is normalized to 1.

The number N$_{i}$ of stars of mass m$_{i}$ is another random variable
whose distribution function follows a Poissonian behavior, with the
value of the IMF at that mass as the only parameter of the
distribution: $dn_{i}=dN_{i}/N_{tot}= \Phi (m_i) dmi$, since the
total number of stars is: $N_{tot} =\int {dN_i }$.

A simple test to check the Poissonian nature of the distribution of
n$_{i}$ is simply the ratio of its variance to its average value as a
function of mass. This ratio should be close to 1.  Cervi\~{n}o et
al. (2001) illustrates in their figure 3 the results of a test with
1000 Montecarlo simulations of clusters with 10$^{3}$ and 10$^{4}$
stars with a Salpeter IMF slope. Despite n$_{i}$ is the ratio of a
Poisson variable with a constant, it is Poissonian distributed within
a 10{\%}.
\begin{figure}[ht]
\psfig{file=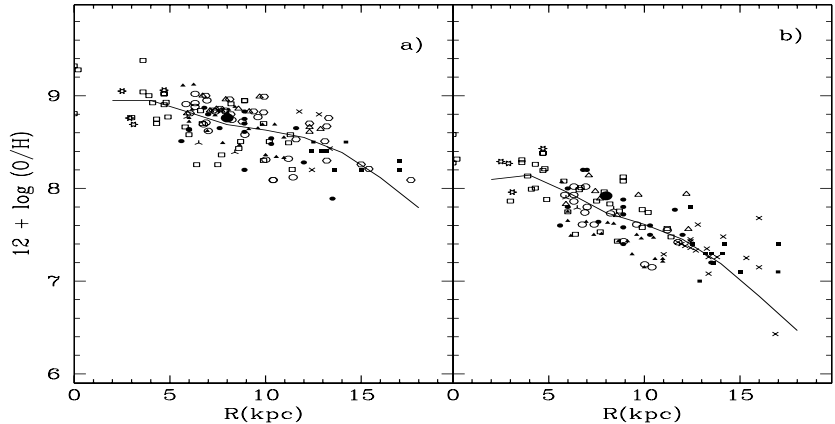,width=12cm,height=6cm,angle=-90}
\caption{Dispersion of O -- panel a)-- and N --panel b)-- abundance
data of the Galaxy taken from different sources}
\label{fig1}
\end{figure}

Once we know that n$_{i}$ can be approximated by a Poisson variable,
it is possible to apply a proper statistical formalism into the code
of our chemical evolution model in order to study the fluctuations which
we may obtain from this stoscastic behavior for the IMF.

\section{The theoretical multiphase chemical evolution model}

Our modeled physical system is a cylindrical sub-galactic zone
centered on the Solar Neighborhood, extending about 1 kpc in the
Galactic plane.  The model used in this work is the code from Ferrini
et al. (1992), improved with recent yields (Gavil\'{a}n et
al. 2002). It computes a production matrix Q$_{ij}$(m), where each
(i,j) element represents the fraction of the star mass initially in
the form of chemical species j, transformed and ejected as chemical
species i, for 15 chemical elements. A matrix is computed for each
stellar mass weigthed by the IMF. The nucleosynthesis production from
SN{\sc I}, SN{\sc II}, and normal stars is taken into account.
We let the model evolves for 13,2 Gyrs, with a 0.01 time step.

The IMF used is taken from Ferrini et al.(1991):

$\Phi (m) = 2.0865m^{ -0.52}x10^{ - [2.07(\log m)^2 + 1.92\log m +
0.73]^{\frac{1}{2}}}$

We will use the two described above methods: a) an analysis of the model
internal error and b) a statistic study on the effects of the IMF
discretization, based on montecarlo models. 

\subsection{Error analysis}
\begin{figure}[ht]
\psfig{file=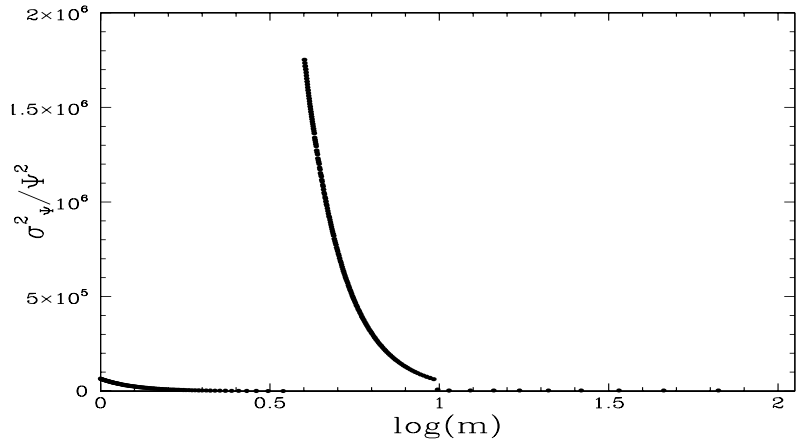,width=11cm,height=6cm,angle=-90}
\caption{Variability of the IMF as a function of mass.}
\label{fig2}
\end{figure}

Considering the Poissonian nature for the IMF, we assume an initial
variance of $\sigma ^2(\Phi ) = \Phi $. Then, we trail the error,
following error propagation rules, within the model in every equation
involved and related to our IMF.  As a result we can obtain a measure
of the variability of our model along the whole mass interval
considered.

We may observe in Fig.~\ref{fig2} that the variability increases
strongly in the mass interval from 4 to 8 M$_{\odot}$.  This is
exactly the interval of mass producing Fe and N. Therefore, in
agreement with this result, our model must show a larger dispersion in
these elements in comparison with other elements as O.

\subsection{Monte Carlo Simulations}

We use an uniformly distributed random numbers generator to generate
from it, and using the accumulated probability distribution, a
succession of numbers with our IMF as probability function. Our aim is
the computation of a complete set of $Q_{i,j}$ Monte Carlo simulations,
where the IMF is randomly different for each of them, with masses from
0.8 up to 100 M$_{\odot}$, divided in 800 intervals with linear
interpolation within them.  We also would like to study the effects of
the discretization caused by rounding in the code. Our aim is to
calculate $\sim 500-1000$ simulations, with $\sim 10^{5}-10^{6}$
stars.  We will repeat our simulations for two different round down
values: 10$^{ - 3}$ and 10$^{ - 6}$. The well calibrated model from
Gavil\'{a}n et al.(2002) will be our base of mean values.

\section{Results} 

\begin{figure}[ht]
\psfig{file=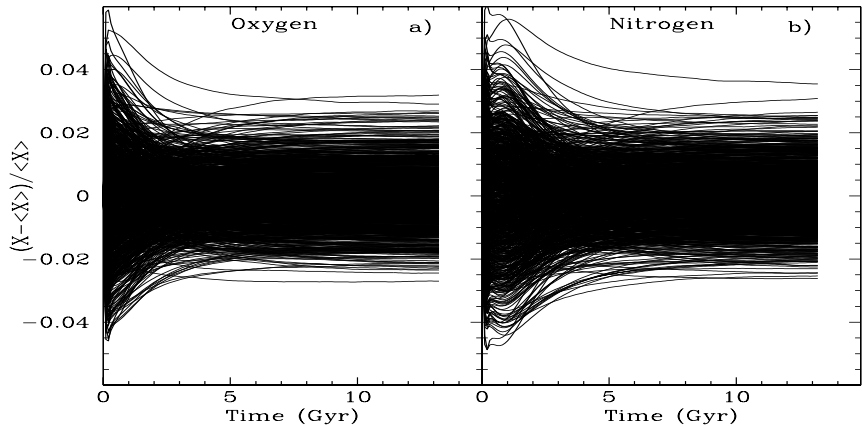,width=12cm,height=6cm,angle=-90}
\caption{Preliminary results of dispersion in O and N abundances.}
\label{fig3}
\end{figure}

The preliminary results have been obtained by the realization of 400
montecarlo simulations for the galaxy disk zone, without having into
account the rounding in the code and with non calibrated models.
Despite of these facts, a dispersion of $\sim 10^{-2}$ already appears
in our O and N abundances. as we may see in Fig.\ref{fig3}.  This fact
seems to indicate that a good percentage of the observed dispersion
could be achieved with the complete set of simulations.

\section{Conclusions}

With this kind of calculations we might estimate the variability and
the errors that can be obtained by the models.  Thus, we could include
into the code the dispersion appearing as a consequence of the
uncertainties in the IMF, and this way we could check if is possible
to get the same dispersion as observed.
It seems, from our preliminary results, that taking into account
the sampling fluctuations from the IMF into the models can reproduce
the observed dispersion.

\end{document}